\documentclass[10pt,english,pre,nofootinbib,reprint, groupedaddress,showpacs,showkeys,preprintnumbers,floatfix]{revtex4-1}
\usepackage[latin9]{inputenc}
\usepackage[a4paper]{geometry}
\usepackage{xcolor}
\usepackage{pdfcolmk}
\usepackage{graphicx}
\usepackage{babel}
\usepackage{mathrsfs}
\usepackage{amsmath}
\usepackage{amssymb}
\usepackage{esint}
\PassOptionsToPackage{normalem}{ulem}
\usepackage{ulem}
\usepackage{dsfont} 
\usepackage[unicode=true,pdfusetitle,
 bookmarks=true,bookmarksnumbered=false,bookmarksopen=false,
 breaklinks=false,pdfborder={0 0 1},backref=false,colorlinks=false]
 {hyperref}

\makeatletter

 \@ifundefined{textcolor}{}
 {%
   \definecolor{BLACK}{gray}{0}
   \definecolor{WHITE}{gray}{1}
   \definecolor{RED}{rgb}{1,0,0}
   \definecolor{GREEN}{rgb}{0,1,0}
   \definecolor{BLUE}{rgb}{0,0,1}
   \definecolor{CYAN}{cmyk}{1,0,0,0}
   \definecolor{MAGENTA}{cmyk}{0,1,0,0}
   \definecolor{YELLOW}{cmyk}{0,0,1,0}
 }

\usepackage{aas_macros}
\usepackage{bbold}

\let\textquotedbl="

\makeatother

\begin{document}

\title{Operator Calculus for Information Field Theory}

\author{Reimar H. Leike, Torsten A. Enßlin}

\affiliation{{\small{}Max-Planck-Institut für Astrophysik, Karl-Schwarzschildstr.~1,
85748 Garching, Germany}\\
Ludwig-Maximilians-Universität München, Geschwister-Scholl-Platz{\small{}~}1,
80539 Munich, Germany}
\begin{abstract}
Signal inference problems with non-Gaussian posteriors can be hard
to tackle. Through using the concept of Gibbs free energy these posteriors
are rephrased as Gaussian posteriors for the price of computing various
expectation values with respect to a Gaussian distribution. We present
a new way of translating these expectation values to a language of
operators which is similar to that in quantum mechanics. This simplifies
many calculations, for instance such involving log-normal priors.
The operator calculus is illustrated by deriving a novel self-calibrating
algorithm which is tested with mock data.
\end{abstract}

\keywords{information theory, information field theory, Bayesian inference,
Gibbs free energy, operator calculus}
\maketitle

\section{Introduction}

Information field theory (IFT) \citep{2009PhRvD..80j5005E} is a Bayesian
formalism for solving field inference problems. Given a prior probability
density and a data model IFT enables us to calculate posterior field
expectation values when data has been measured. A common way to summarize
the posterior is in terms of an estimate of the signal posterior mean
and its variance. One way to obtain an approximation to those is provided by 
the Gibbs free energy  method \citep{opper2001advanced, 2010PhRvE..82e1112E}, also called ``variational Bayes'' 
or ``mean field approximation''. The minimum of the Gibbs free energy
is an estimate of the posterior mean. The curvature of this minimum
encodes the posterior covariance. The involved mathematical expressions
contain Gaussian field averages over many functions which can become
very difficult to evaluate especially in case of log-normal signal
distributions, interactive Hamiltonians, or non-linear responses. 

We present a new way to calculate such Gaussian integrals; by translating
them to operator calculations and using well-known formulas from differential
geometry we are able to handle them efficiently. This is a general
technique for calculating expectation values over Gaussian distributions
which could have implications to other contexts as well. We chose
to apply it to the Gibbs formalism because on the one hand Gibbs free
energy inference is a very general tool for tackling inference problems
and on the other hand because in this context the introduced operator
formalism proves to be exceedingly useful.

In Sect.\,\ref{sec:Gibbs-Free-Energy} we give a short review of
Ref.\,\citep{2010PhRvE..82e1112E}, introducing the reader to the concept
of Gibbs free energy inference and explaining its advantages and challenges.
In Sect.\,\ref{sec:Self-Calibrating} we introduce a typical problem set
of image reconstruction as an example.
In Sect.\,\ref{sec:Formulating-Gaussian-Averages} we translate expectation
values over a Gaussian distribution into the language of operators.
We then show how to leverage the power of our operator calculus with a certain set of
algebraic tools in Sect.\,\ref{Calculating-Gaussian-Expectation}.
The algorithm that is derived using these algebraic tools is then implemented
and tested for mock data. The results are discussed in Sect.\,\ref{sec:Numerical-Example}.
We conclude in Sect.\,\ref{sec:Conclusion}.
The derivation of the algorithm for the image reconstruction problem introduced earlier
is in appendix\,\ref{sec:Facilitating-Calculations-with}.

\section{Gibbs Free Energy Inference\label{sec:Gibbs-Free-Energy}}

To give a better understanding of the benefit we get from the operator
calculus to be introduced in chapter \ref{sec:Formulating-Gaussian-Averages}, we first give a brief introduction
to the Gibbs formalism \citep{2010PhRvE..82e1112E}. 

In signal reconstruction we try to infer a signal $s$ when given
the data
\begin{equation}
d=r(s,n)
\end{equation}
for a given response operator $r$ and measurement noise $n$. Here
\begin{align}
s:X & \rightarrow\mathbb{C}\nonumber \\
x & \mapsto s(x)=s_{x}
\end{align}
 is a field over some measure space $X$. In order to infer the signal,
we use the posterior probability density
\begin{equation}
P(s|d)=\frac{P(d|s)P(s)}{P(d)}=\frac{P(d|s)P(s)}{\int\mbox{d}s\,P(d|s)P(s)}\ .
\end{equation}
For more complicated problems, this posterior probability density
is often not accessible because integrations like that in the denominator
might not be analytically solvable. In most situations we do however
have access to the so called information Hamiltonian $H(s,d)=-\mbox{ln}(P(s))-\mbox{ln}(P(d|s))$
which contains all information available on the signal $s$. We can
ignore additive constants in the Hamiltonian that depend only on $d$
since they cancel when we reconstruct $P(s|d)$ from the Hamiltonian,
\begin{align}
P(s|d) & =\frac{e^{-H(s,d)}}{Z(d)}\\
\mbox{where }Z(d) & =\int\mbox{d}s\,e^{-H(s,d)}
\end{align}
is the partition function. Thus we try to infer $s$ using an estimator
that only uses the Hamiltonian. A simple way is by minimizing $H(s,d)$,
but this yields suboptimal results for asymmetric posterior distributions
as well as unsatisfying error estimates for posterior distributions
that deviate strongly from Gaussianity, see e.g. \citep{2010arXiv1002.2928Ep}.
There exist other estimators, but our formalism of operator calculus
is most suited for the Gibbs free energy method and thus we will concentrate
on this approach. In the Gibbs formalism we approximate the posterior
by a Gaussian distribution 
\begin{equation}
\tilde{P}(s|d)=\mathscr{G}(s-m,D)=\frac{e^{\frac{1}{2}(s-m)^{\dagger}D^{-1}(s-m)}}{\left|2\pi D\right|^{\frac{1}{2}}}
\end{equation}
with mean $m$ and covariance $D$ that depend on the data in a way
still to be found. Here we use the scalar product of fields
\begin{equation}
a^{\dagger}b=\int\text{d}x\,a_{x}^{*}b_{x}
\end{equation}
 with $*$ denoting complex conjugation. For a Gaussian posterior,
calculating the partition function is possible (see Sect.\,II E of Ref.\,\citep{2010PhRvE..82e1112E}). 
This approach is also known in the literature under the names of
``variational Bayes'' and ``mean field approximation'' but we will 
call it Gibbs free energy because it was historically the earliest of the three concepts.

Almost every scientific result is given as a pair of estimate and
standard deviation. Most people assume Gaussian statistics when seeing
a result like that, consistent with its maximum entropy translation
into a probability density function. Thus approximating the posterior
by a Gaussian is basically common practice and information theoretically
supported \citep{2003prth.book.....J}. While doing so, we might as
well try to invent the least amount of information through the approximation.
In order to obtain $m$ and $D$ we therefore minimize the Kullback-Leibler
divergence \citep{Kullback1951}:
\begin{align}
d_{KL}(\tilde{P},P)= & \int\text{d}s\,\tilde{P}(s|d)\text{ln}\left(\frac{\tilde{P}(s|d)}{P(s|d)}\right)\nonumber \\
= & \left\langle \text{ln}\left(\frac{\tilde{P}(s|d)}{P(s|d)}\right)\right\rangle _{\mathscr{G}(s-m,D)}\nonumber \\
= & \left\langle \text{ln}\left(\tilde{P}(s|d)\right)\right\rangle _{\mathscr{G}(s-m,D)}\nonumber \\
 & +\left\langle H(s|d)\right\rangle _{\mathscr{G}(s-m,D)}\label{eq:Gibbs-formula}
\end{align}
The Kullback-Leibler divergence is therefore up to the irrelevant
constant $\mbox{ln}\left(Z(d)\right)$ equal to the Gibbs free energy
$G(m,D)=U-TS$ at temperature $T=1$ with the Shannon entropy \citep{Shannon1948}
\begin{equation}
S=-\left\langle \text{ln}\left(\tilde{P}(s|d)\right)\right\rangle _{\mathscr{G}(s-m,D)}
\end{equation}
 and the internal energy 
\begin{equation}
U=\left\langle H(s,d)\right\rangle _{\mathscr{G}(s-m,D)}\ .
\end{equation}
The posterior mean $m$ is now given within our approximation by the
minimum of the Gibbs free energy
\begin{equation}
m:\;\frac{\delta G}{\delta m}=0\ .
\end{equation}
From a Kullback-Leibler divergence perspective, the posterior uncertainty
dispersion is given by
\begin{equation}
D:\;\frac{\delta G}{\delta D}=0\ .
\end{equation}
Derivatives with respect to operators can be difficult to handle.
Fortunately the thermodynamical relation 
\begin{equation}
D=\left(\frac{\delta^{2}G}{\delta m\delta m^{\dagger}}\right)^{-1}\ ,
\end{equation}
which holds at the minimum of the Gibbs free energy, 
requires only derivatives with respect to the mean field $m$.

Note that following the Gibbs free energy approach we only need to
calculate expectation values over a Gaussian distribution and instead
of the full posterior $P(s|d)$ it suffices to know the Hamiltonian
$H(s,d)$ of the joint probability of data and signal.

\section{A Self-Calibrating System as Example\label{sec:Self-Calibrating}}

Suppose now we have a measurement scenario where a real field $a$
contributes to the data via being exponentiated 
\begin{align}
d & =re^{a}+n\\
P(a) & =\mathscr{G}(a,A)\label{eq:prior-signal}\ .
\end{align}
This corresponds to a linear data model with response operator $r$
and a log-normal prior which is the natural prior for strictly positive
signals that vary over orders of magnitude. For example the galaxy
densities in the cosmos show roughly log-normal distributions as was supported
empirically \citep{1934ApJ....79....8H,2009MNRAS.400..183K} and theoretically
\citep{Layzer1956,1991MNRAS.248....1C,1995MNRAS.277..933S,Kayo2001,2001PASP..113.1009V,2009ApJ...698L..90N}.
Suppose additionally that the response operator is unknown and we
have an independent Gaussian prior for it
\begin{equation}
P(r)=\mathscr{G}(r,R)\label{eq:prior-response}\ .
\end{equation}
Now we are actually dealing with a self-calibration problem, as the
unknown instrument response $r$ has to be inferred from the unknown
signal observation. These are notoriously hard, see \citep{2015PhRvE..91a3311D}
or \citep{2014PhRvE..90d3301E}. 

We define a joint signal vector 
\begin{align}
s &=\left(\begin{array}{c}
r\\
a
\end{array}\right)\\
P(s) & =\mathscr{G}(s,S)=\mathscr{G}\left(s,\left(\begin{array}{cc}
R & 0\\
0 & A
\end{array}\right)\right)\label{eq:Definition D}
\end{align}
for all quantities we would like to infer.

In the simple case of Gaussian additive noise $P(n)=\mathscr{G}(n,N)$
we get as Hamiltonian 
\begin{align}
H(d,s) &= H(s)+H(d|s)\nonumber \\
&= \frac{1}{2}s^{\dagger}S^{-1}s\nonumber \\
 &\quad +\frac{1}{2}\left(d-re^{a}\right)^{\dagger}N^{-1}\left(d-re^{a}\right)\ .\label{eq:Hamiltonian}
\end{align}
The Hamiltonian thus contains the interacting signal terms $d^{\dagger}N^{-1}re^{a}$
and $(re^{a})^{\dagger}N^{-1}re^{a}$ for which the expectation value over
the generic Gaussian distribution $\mathscr{G}(s-m,D)$ has to be taken
to calculate the Gibbs free energy. Although this can be done by hand, calculations
can get very tedious and require a lot of time. We will be able to
handle them quite nicely with our operator formalism in appendix\,\ref{sec:Facilitating-Calculations-with}.

In case of a non-linear response or a signal depended noise model
we get even more exponentials and potentially additional factors of
polynomials in $s$. 

Now that we have seen a typical problem set, let us proceed by introducing
the tools to translate expectation values over Gaussian distributions
to operator action.

\section{Formulating Gaussian Averages in Operator Calculus\label{sec:Formulating-Gaussian-Averages}}

In this section we are concerned with the task of calculating the
expectation value 
\begin{equation}
\left<f(s)\right>_{\mathscr{G}(s-m,D)}
\end{equation}
for a Gaussian distribution in $s$ with mean $m$ and covariance
$D$. 

Let us start with the much more simple task of calculating $\left<s\right>_{\mathscr{G}(s-m,D)}$.
We let us guide by a calculation trick from statistical physics where
a lot of expectation values are calculated by taking different derivatives
of the partition sum and thus try to obtain $s$ by taking the derivative
of $\mathscr{G}(s-m,D)$: 
\begin{align}
&\frac{\delta}{\delta m}\mathscr{G}(s-m,D)  =D^{-1}(s-m)\mathscr{G}(s-m,D)\nonumber \\
&\Rightarrow(D\frac{\delta}{\delta m}+m)\mathscr{G}(s-m,D)  =s\,\mathscr{G}(s-m,D)
\end{align}
 Thus we have
\begin{equation}
\left<s\right>_{\mathscr{G}(s-m,D)}=\left<D\frac{\delta}{\delta m}+m\right>_{\mathscr{G}(s-m,D)}\ .
\end{equation}
Here the linear operator $D\frac{\delta}{\delta m}+m$
does not depend on $s$, so one may pull it out of the expectation
value:
\begin{equation}
\left<s\right>_{\mathscr{G}(s-m,D)}=\left(D\frac{\delta}{\delta m}+m\right)\left<1\right>_{\mathscr{G}(s-m,D)}=m
\end{equation}
This is not a surprising result. However, it is remarkable that this
works for any moment of the Gaussian
\begin{equation}
\left<s^{n}\right>_{\mathscr{G}(s-m,D)}=\left(D\frac{\delta}{\delta m}+m\right)^{n}1\ .
\end{equation}
We call $\Phi:=D\frac{\delta}{\delta m}+m$ the
$s$-operator. Let's look at the expectation value of an arbitrary
analytical function $f$. By definition, an analytical function can
be expanded locally in a point $s_0$ in a series $f(s)=\sum_{n=0}^{\infty}\Lambda_{n}(s-s_0)^{n}$
that has a positive convergence radius. We use a short notation for
the Taylor-Fréchet expansion of the function $f$,
\begin{align}
f(s)=&\sum_{n=0}^{\infty}\Lambda_{n}\,(s-s_0)^{n}\nonumber \\
=&\sum_{n=0}^{\infty}\int\textnormal{d}x_{1}\dots\int\textnormal{d}x_{n}\nonumber \\
&\qquad \Lambda_{n}(x_{1},\dots,x_{n}) (s-s_0)_{x_{1}}\dots(s-s_0)_{x_{n}}\label{eq:Taylor-Frechet}
\end{align}
and calculate
\begin{align}
 &\left<f(s)\right>_{\mathscr{G}(s-m,D)} =\nonumber\\
&=\sum_{n=0}^{\infty}\Lambda_{n}\left<(s-s_0)^{n}\right>_{\mathscr{G}(s-m,D)}\nonumber \\
 & =\sum_{n=0}^{\infty}\Lambda_{n}\sum_{i=0}^{n}\left(\begin{array}{c}
n\\
i
\end{array}\right)\left<s^{i}\left(-s_0\right)^{n-i}\right>_{\mathscr{G}(s-m,D)}\nonumber \\
 & =\sum_{n=0}^{\infty}\Lambda_{n}\sum_{i=0}^{n}\left(\begin{array}{c}
n\\
i
\end{array}\right)\left<\Phi^{i}\left(-s_0\right)^{n-i}\right>_{\mathscr{G}(s-m,D)}\nonumber \\
 & =\sum_{n=0}^{\infty}\Lambda_{n}\left<(\Phi-s_0)^{n}\right>_{\mathscr{G}(s-m,D)}\nonumber \\
 & =\sum_{n=0}^{\infty}\Lambda_{n}\left(\Phi-s_0\right)^{n}1=f(\Phi)1\ .
\end{align}
Thus instead of calculating the expectation value of $f(s)$ with
respect to a Gaussian distribution we can let the operator $f(\Phi)$
act on $1$.

When dealing with complex numbers we have to treat $s$ and $s^{*}$
separately and replace them with $\Phi^{\prime}:=2D\frac{\delta}{\delta m^{*}}+m$
and $\Phi^{\prime*}:=2D\frac{\delta}{\delta m}+m^{*}$
respectively. These two operators commute $[\Phi^{\prime},\Phi^{\prime*}]=0$
and calculations thus follow a similar line for complex fields.

\section{Calculating Gaussian Expectation Values Algebraically\label{Calculating-Gaussian-Expectation}}

In order to highlight the benefit of this reformulation of integrations to operator actions, we
introduce the reader to certain useful algebraic tools and show how to apply them. 
The first step to all calculations is to separate 
\begin{align}
\Phi_x=\int \text{d}y\,D_{xy}\frac{\delta}{\delta m_y}+m_x=c^{x}+b^{x}\ .
\end{align}
We call $b^{x} = m_x$ the creation operator and $c^{x} = \int \text{d}y\,D_{xy}\frac{\delta}{\delta m_y}$ 
the annihilation operator.
Our goal is to get the annihilation operators to the right hand side because
they cancel 
\begin{align}
c^{x}1=\int\mbox{d}t\,D_{xt}\frac{\delta}{\delta m_{t}}1=0\ .
\end{align}
To achieve this we use the commutation relations of the creation and annihilation operators 
\begin{align}
[b^{x},b^{y}]=[c^{x},c^{y}] & =0\\
[c^{x},b^{y}] & =D_{xy}\ .
\end{align}
How exactly we bring the annihilation part to the right side differs for
different classes of functions.
For polynomials we can simply use distributivity of multiplication
\begin{align}
\Phi_x\Phi_y&=(b^x+c^x)(b^y+c^y)\nonumber\\
&=b^xb^y+b^xc^y+c^xb^y+c^xc^y
\end{align}
and then apply the commutation relations to obtain
\begin{align}
\Phi_x\Phi_y&=b^xb^y+2b^xc^y+[c^x,b^y]+c^xc^y\nonumber\\
&=b^xb^y+2b^xc^y+D_{xy}+c^xc^y\nonumber\\
\Rightarrow\Phi_x\Phi_y1&=m_xm_y+D_{xy}\ .
\end{align}
We can separate creation and annihilation parts for exponential functions by
making use of the Baker-Campbell-Hausdorff (BCH) formula \citep{campbell1897law} 
\begin{equation}
e^{b^{x}+c^{y}+\frac{1}{2}[b^{x},c^{y}]}=e^{b^{x}}e^{c^{y}}\ .
\end{equation}
Thereby, we can omit further iterations of the commutator that appear in the full BCH formula because
$[b^{x},c^{y}]=-D_{xy}$ is central in the
algebra of linear operators on functions of $m$, i.e. it commutes
with $c^{y}$ and $b^{x}$. Applying this yields
\begin{align}
e^{\Phi_x} &= e^{b^{x}+c^{x}}\nonumber\\
&=e^{-\frac{1}{2}[b^{x},c^{x}]}e^{b^{x}}e^{c^{x}}\nonumber\\
&=e^{\frac{1}{2}D_{xx}}e^{b^{x}}e^{c^{x}}
\end{align}
Thus for certain functions $f(\Phi)$
we are able to separate the annihilation part and the creation part of $\Phi$
\begin{align}
 f(\Phi) = \sum_i f^{b}_i(b)f^{c}_i(c)
\end{align}
using algebraic tools. One major advantage of using that approach instead of calculating the expectation value directly is that now
calculating the expectation value of the product of two functions $\left\langle f(s)g(s)\right\rangle_{\mathscr{G}(s-m,D)}$ simply amounts to calculating the commutator
\begin{align}
  \left\langle f(s)g(s)\right\rangle_{\mathscr{G}(s-m,D)} =& f(\Phi)g(\Phi)1 \nonumber\\
 =& \sum_i f^{b}_i(b)f^{c}_i(c)\sum_j g^{b}_i(b)g^{c}_i(c)1\nonumber\\
 =& \sum_{i,j} f^{b}_i(m)g^{b}_i(m)f^{c}_i(0)g^{c}_i(0)1\nonumber\\
  &+ \sum_{i,j} f^{b}_i(b)\left[f^{c}_i(c), g^{b}_i(b)\right]g^{b}_i(0)1\nonumber\\
 =&  \left\langle f(s)\right\rangle_{\mathscr{G}(s-m,D)} \left\langle g(s)\right\rangle_{\mathscr{G}(s-m,D)}\nonumber\\
 & + \sum_{i,j} f^{b}_i(b)\left[f^{c}_i(c), g^{b}_i(b)\right]g^{c}_i(0)1
\end{align}
of the two involved functions. 

We can calculate those commutators using algebraic tools.
For example to exchange  $c^{x}$ and $e^{b^{y}}$ we use the fact that
$[c^{x},\_]$ has the algebraic properties of a derivation, meaning
it is linear and obeys the product rule
\begin{equation}
[c^{x},ab]=[c^{x},a]b+a[c^{x},b]\ .
\end{equation}
Thus 
\begin{align}
[c^{x},e^{b^{y}}] & =\sum_{n=0}^{\infty}\frac{[c^{x},\left(b^{y}\right)^{n}]}{n!}\nonumber \\
 & =\sum_{n=0}^{\infty}\frac{n\left(b^{y}\right)^{n-1}[c^{x},b^{y}]}{n!}\nonumber \\
 & =e^{b^{y}}[c^{x},b^{y}]\nonumber \\
 & =D_{xy}e^{b_{y}}\ .\label{eq:Commutator-Monomial-Exponential}
\end{align}
We can calculate the commutator of two exponential functions using the BCH-formula twice:
\begin{align}
[e^{c^{x}},e^{b^{y}}] &= e^{c^{x}}e^{b^{y}} - e^{b^{y}} e^{c^{x}}\nonumber\\
&= e^{b^{y}+c^{x}-\frac{1}{2}[b^{y},c^{x}]} - e^{b^{y}}e^{c^{x}} \nonumber\\
&= e^{b^{y}}e^{c^{x}}e^{-[b^{y},c^{x}]}- e^{b^{y}}e^{c^{x}}\nonumber\\
&= e^{b^{y}}e^{c^{x}}\left(e^{D_{xy}}-1\right)\label{eq:Commutator-Exponential-Exponential}
\end{align}
If we just want to exchange the position of these exponentials the formula (\ref{eq:Commutator-Exponential-Exponential}) simplifies to
\begin{equation}
e^{c^{x}}e^{b^{y}} = e^{b^{y}}e^{c^{x}}e^{D_{xy}} \ .
\end{equation}
Having aggregated these tools, calculating the Gibbs free energy of the self-calibration problem introduced in Sec.\,\ref{sec:Self-Calibrating}
is straight forward. This calculation is done in appendix \ref{sec:Facilitating-Calculations-with}.

\begin{figure*}[t]
  \begin{minipage}[t]{\columnwidth}
   \includegraphics[width=\textwidth]{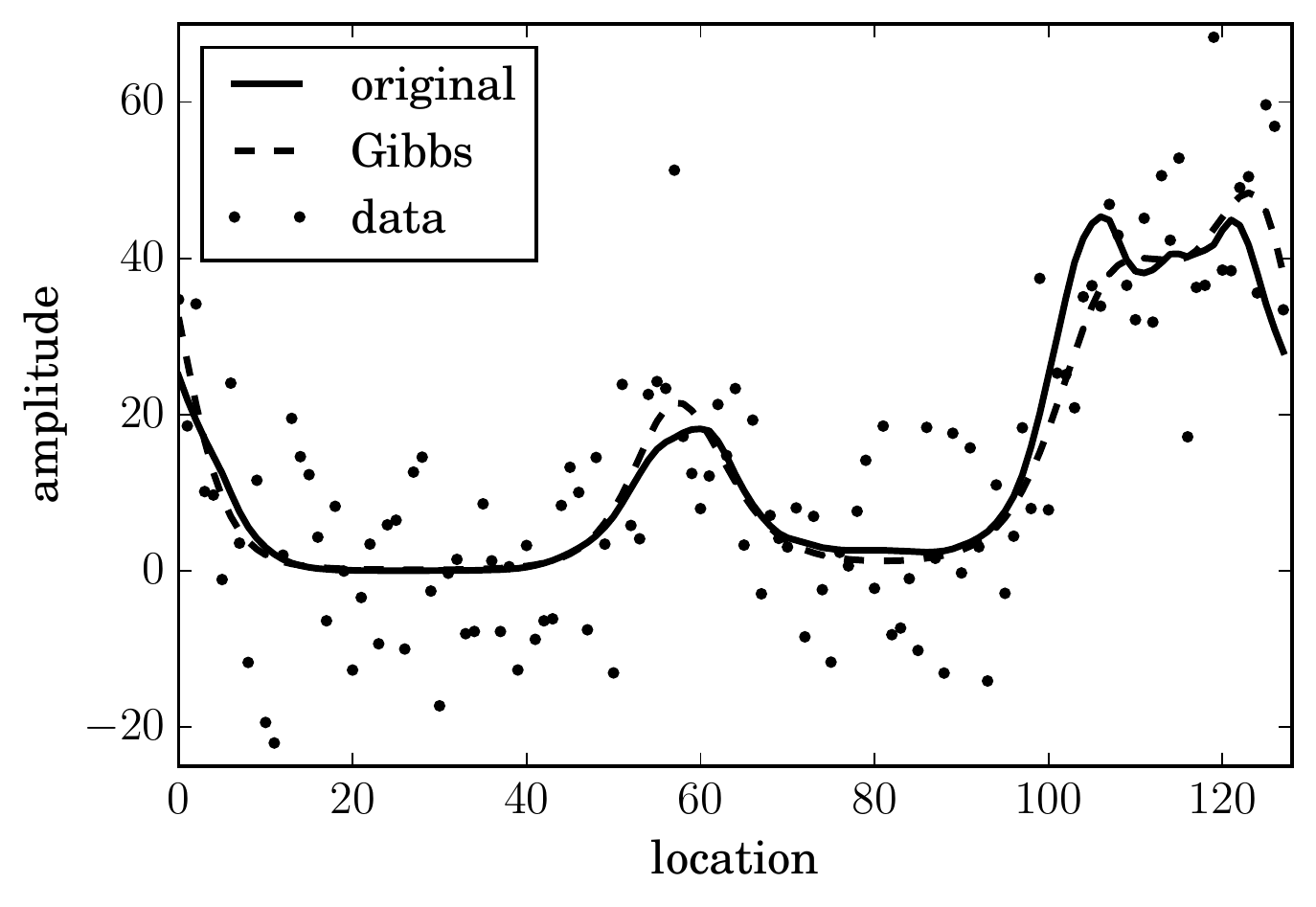}
   \caption{Original signal response (solid line), expected value of the signal response given via the Gibbs estimate (dashed line)
and data that was sampled from the prior (points). The expected value of the signal response was computed with the formula $(r_0+m_r+D_{ra})e^{m_a+\frac{1}{2}\widehat{D_{aa}}}$. }
   \label{fig:both-estimates}
  \end{minipage}
  \hfill
  \begin{minipage}[t]{\columnwidth}
    \includegraphics[width=0.985\textwidth]{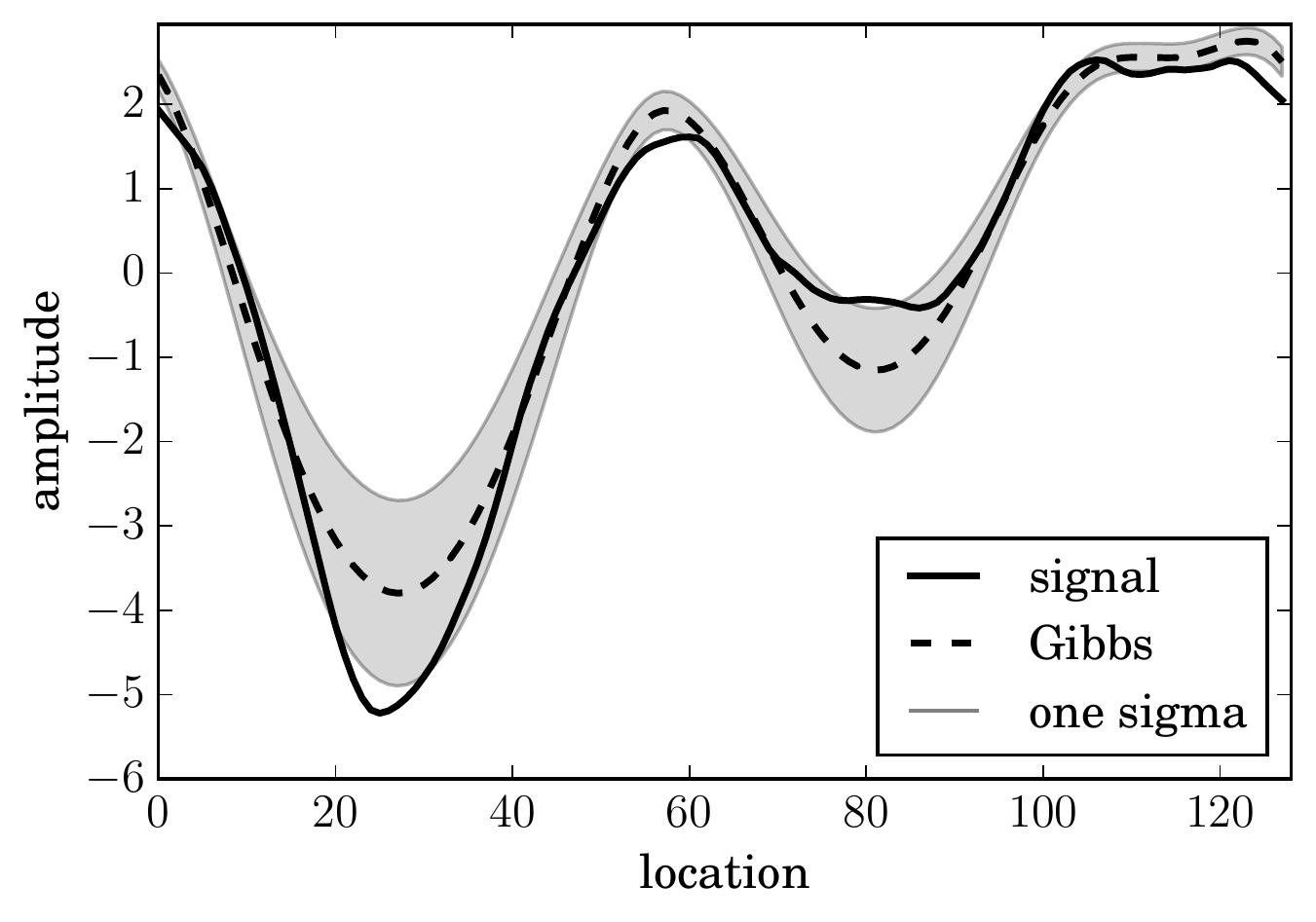}
    \caption{  
    The signal reconstruction $m_a$ (dashed line) with position dependent uncertainty range (shaded area). This uncertainty is given by the square root of the diagonal
    of the covariance matrix $\widehat{D_{aa}}$.
    Notice that due to the log-normal nature of the problem, higher signal values are reconstructed more accurately.}
    \label{fig:gibbs-result}
  \end{minipage}
\end{figure*}

\section{A Numerical Example \label{sec:Numerical-Example}}

To conclude this paper we implemented and verified the derived algorithm.
For our implementation we use the slightly altered data model 
\begin{equation}
d=\left(r+r_0\right)\mathds{1}e^{a}+n
\end{equation}
where $\mathds{1}$ denotes the identity matrix and $r$ is just a
scalar. This simplified model has the advantage of being easier to implement because it is less degenerate and its results are easier to visualize. The constant $r_0$ encodes
that we usually have some rough idea about the typical response of 
our instrument.
We assume a Gaussian noise distribution
$$
P(n) = \mathscr{G}(n,\sigma_N^2\delta_{ij})
$$
that has a scalar covariance $\sigma_N^2$ and also a Gaussian prior distribution for $a$ and $r$ as in equations (\ref{eq:prior-signal}) and (\ref{eq:prior-response}).
The only difference is that
the covariance matrix for $r$ is now just a constant. We take the signal prior covariance $A$ to be diagonal in Fourier space with known power spectrum.
Analogously to the derivation in appendix\,\ref{sec:Facilitating-Calculations-with} we arrive at
\begin{align}
 &G(m,D)=-\frac{1}{2}\mbox{tr}\left(1+\mbox{ln}\left(2\pi D\right)\right)\nonumber\\
 & +\frac{1}{2}m^{\dagger}S^{-1}m+\frac{1}{2}\mbox{tr}\left(S^{-1}D\right)\nonumber \\
 & -\int\mbox{d}i\,\left(\frac{d^{\dagger}}{\sigma_N^2}\right)_{i}\left(m_{r}+r_0+\left(D_{ra}\right)_{i}\right)\left(e^{m_{a}+\frac{1}{2}\widehat{D_{aa}}}\right)_{i}\nonumber \\
 & +\frac{1}{2}\int\mbox{d}i\, \left(e^{2m_{a}+2\widehat{D_{aa}}}\right)_{i}\nonumber\\
 & \quad\left(D_{rr}+\left(m_{r}+r_0+2\left(D_{ra}\right)_{i}\right)^2\right)\ .
\end{align}
Taking the derivative by $m$ we arrive at the gradient in signal direction
\begin{align}
 & \frac{\delta G(m,D)}{\delta \left(m_a\right)_i} = A^{-1}m_a\nonumber\\
 & -\left(\frac{d^{\dagger}}{\sigma_N^2}\right)_{i}\left(m_{r}+r_0+\left(D_{ra}\right)_{i}\right)\left(e^{m_{a}+\frac{1}{2}\widehat{D_{aa}}}\right)_{i}\nonumber \\
 & \quad+\left(e^{2m_{a}+2\widehat{D_{aa}}}\right)_{i}\left(D_{rr}+\left(m_{r}+r_0+2\left(D_{ra}\right)_{i}\right)^2\right)
\end{align}
and in the response factor direction
\begin{align}
 & \frac{\delta G(m,D)}{\delta m_r}= \frac{m_r}{R}-\int\mbox{d}i\left(\frac{d^{\dagger}}{\sigma_N^2}\right)_{i}\left(e^{m_{a}+\frac{1}{2}\widehat{D_{aa}}}\right)_{i}\nonumber \\
 & \quad+\int\mbox{d}i\left(e^{2m_{a}+2\widehat{D_{aa}}}\right)_{i}\left(m_{r}+r_0+2\left(D_{ra}\right)_{i}\right)                  \ ,
\end{align}
respectively.
Taking the derivative again we arrive at the Hessian matrix
\begin{align}
 & \frac{\delta^2 G(m,D)}{\delta (m_a)_i\delta (m_a^\dagger)_j} = A^{-1}_{ij}\nonumber\\
 & -\delta_{ij}\left(\frac{d^{\dagger}}{\sigma_N^2}\right)_{i}\left(m_{r}+r_0+\left(D_{ra}\right)_{i}\right)\left(e^{m_{a}+\frac{1}{2}\widehat{D_{aa}}}\right)_{i}\nonumber \\
 & +2\delta_{ij}\left(e^{2m_{a}+2\widehat{D_{aa}}}\right)_{i}\left(D_{rr}+\left(m_{r}+r_0+2\left(D_{ra}\right)_{i}\right)^2\right)
\end{align}
\begin{align}
 &  \frac{\delta^2 G(m,D)}{\delta m_r\delta (m_a)_i}=-\left(\frac{d^{\dagger}}{\sigma_N^2}\right)_{i}\left(e^{m_{a}+\frac{1}{2}\widehat{D_{aa}}}\right)_{i}\nonumber\\
 & +2\left(e^{2m_{a}+2\widehat{D_{aa}}}\right)_{i}\left(m_{r}+r_0+2\left(D_{ra}\right)_{i}\right)
\end{align}
\begin{align}
 & \frac{\delta^2 G(m,D)}{\delta^2 m_r}= R+\int\mbox{d}i\left(e^{2m_{a}+2\widehat{D_{aa}}}\right)_{i}\ .
\end{align}
Using these in a Newton scheme to find the minimum, we arrive at an algorithm that we implemented. 
In Fig.\,\ref{fig:both-estimates} we show mock data that was generated by sampling from the prior in comparison to the projected signal response $(r+r_0)e^a$ and 
the expected value of the signal response as was computed by the Gibbs
reconstruction algorithm.
Notice that the signal response varies over a few orders of magnitudes due to its log-normal nature. One can also see that the signal has strong spatial 
correlations that were encoded into the prior. 
Fig.\,\ref{fig:gibbs-result} shows the reconstruction of the signal we get from our algorithm. These results only deviate slightly
from the results we get when using a maximum a posteriori (MAP) estimator
because the posterior is still relatively near to a Gaussian. The response factor $r+r_0$ was sampled to be $3.62$, the corresponding Gibbs estimate is $m_r+r_0 = 3.11 \pm 0.41$ and the MAP 
estimate of it is $\left(m_{\text{MAP}}\right)_r+r_0 = 3.15 \pm 0.46$. Thus both deviate about one sigma from the actual value. In this case, the Gibbs result is comparable to the 
MAP estimator. However, this inference problem was chosen to demonstrate how the operator formalism works and not to highlight the differences in the performance
of the Gibbs estimator with respect to that of the MAP estimator.

\section{Conclusion\label{sec:Conclusion}}

With the help of the Gibbs free energy one can easily write
down expressions for the posterior mean and covariance. Using the
operator formalism introduced in this paper we formulated expectation
values as operators acting on $1$ which eliminates the need to calculate
an integral over the Hamiltonian and the Gaussian distribution. This
process of translating the expectation value to operators works generally,
albeit possibly entailing algebraic complexity. For expectation values
over products of exponential functions and polynomials which are typical
for problems with log-normal statistics, we aggregated a collection
of algebraic tools that enable us to nevertheless calculate them in
a few lines of straight forward calculation. We demonstrated their
usage by applying our operator calculus to a signal inference problem
with log-normal prior and unknown but linear response operator for
which we worked out all the occurring terms with regard to the posterior
mean. The resulting algorithm was implemented and found to be working for mock data.

Future research might be directed towards finding analogies to the
BCH formula for function classes other than the exponential function
which will allow us to apply the operator formalism to an even broader
range of problems.

\section{Acknowledgments}

We acknowledge valuable discussions and comments on the manuscript
by Maksim Greiner, Theo Steininger, Jakob Knollmüller and Sebastian
Hutschenreuter.

\bibliographystyle{apsrev4-1}

\bibliography{ift}

\begin{thebibliography}{18}%
\makeatletter
\providecommand \@ifxundefined [1]{%
 \@ifx{#1\undefined}
}%
\providecommand \@ifnum [1]{%
 \ifnum #1\expandafter \@firstoftwo
 \else \expandafter \@secondoftwo
 \fi
}%
\providecommand \@ifx [1]{%
 \ifx #1\expandafter \@firstoftwo
 \else \expandafter \@secondoftwo
 \fi
}%
\providecommand \natexlab [1]{#1}%
\providecommand \enquote  [1]{``#1''}%
\providecommand \bibnamefont  [1]{#1}%
\providecommand \bibfnamefont [1]{#1}%
\providecommand \citenamefont [1]{#1}%
\providecommand \href@noop [0]{\@secondoftwo}%
\providecommand \href [0]{\begingroup \@sanitize@url \@href}%
\providecommand \@href[1]{\@@startlink{#1}\@@href}%
\providecommand \@@href[1]{\endgroup#1\@@endlink}%
\providecommand \@sanitize@url [0]{\catcode `\\12\catcode `\$12\catcode
  `\&12\catcode `\#12\catcode `\^12\catcode `\_12\catcode `\%12\relax}%
\providecommand \@@startlink[1]{}%
\providecommand \@@endlink[0]{}%
\providecommand \url  [0]{\begingroup\@sanitize@url \@url }%
\providecommand \@url [1]{\endgroup\@href {#1}{\urlprefix }}%
\providecommand \urlprefix  [0]{URL }%
\providecommand \Eprint [0]{\href }%
\providecommand \doibase [0]{http://dx.doi.org/}%
\providecommand \selectlanguage [0]{\@gobble}%
\providecommand \bibinfo  [0]{\@secondoftwo}%
\providecommand \bibfield  [0]{\@secondoftwo}%
\providecommand \translation [1]{[#1]}%
\providecommand \BibitemOpen [0]{}%
\providecommand \bibitemStop [0]{}%
\providecommand \bibitemNoStop [0]{.\EOS\space}%
\providecommand \EOS [0]{\spacefactor3000\relax}%
\providecommand \BibitemShut  [1]{\csname bibitem#1\endcsname}%
\let\auto@bib@innerbib\@empty
\bibitem [{\citenamefont {{En{\ss}lin}}\ \emph {et~al.}(2009)\citenamefont
  {{En{\ss}lin}}, \citenamefont {{Frommert}},\ and\ \citenamefont
  {{Kitaura}}}]{2009PhRvD..80j5005E}%
  \BibitemOpen
  \bibfield  {author} {\bibinfo {author} {\bibfnamefont {T.~A.}\ \bibnamefont
  {{En{\ss}lin}}}, \bibinfo {author} {\bibfnamefont {M.}~\bibnamefont
  {{Frommert}}}, \ and\ \bibinfo {author} {\bibfnamefont {F.~S.}\ \bibnamefont
  {{Kitaura}}},\ }\href {\doibase 10.1103/PhysRevD.80.105005} {\bibfield
  {journal} {\bibinfo  {journal} {\prd}\ }\textbf {\bibinfo {volume} {80}},\
  \bibinfo {pages} {105005} (\bibinfo {year} {2009})},\ \Eprint
  {http://arxiv.org/abs/0806.3474} {arXiv:0806.3474} \BibitemShut {NoStop}%
\bibitem [{\citenamefont {Opper}\ and\ \citenamefont
  {Saad}(2001)}]{opper2001advanced}%
  \BibitemOpen
  \bibfield  {author} {\bibinfo {author} {\bibfnamefont {M.}~\bibnamefont
  {Opper}}\ and\ \bibinfo {author} {\bibfnamefont {D.}~\bibnamefont {Saad}},\
  }\href@noop {} {\emph {\bibinfo {title} {Advanced mean field methods: Theory
  and practice}}}\ (\bibinfo  {publisher} {MIT press},\ \bibinfo {year}
  {2001})\BibitemShut {NoStop}%
\bibitem [{\citenamefont {{En{\ss}lin}}\ and\ \citenamefont
  {{Weig}}(2010)}]{2010PhRvE..82e1112E}%
  \BibitemOpen
  \bibfield  {author} {\bibinfo {author} {\bibfnamefont {T.~A.}\ \bibnamefont
  {{En{\ss}lin}}}\ and\ \bibinfo {author} {\bibfnamefont {C.}~\bibnamefont
  {{Weig}}},\ }\href {\doibase 10.1103/PhysRevE.82.051112} {\bibfield
  {journal} {\bibinfo  {journal} {\pre}\ }\textbf {\bibinfo {volume} {82}},\
  \bibinfo {eid} {051112} (\bibinfo {year} {2010})},\ \Eprint
  {http://arxiv.org/abs/1004.2868} {arXiv:1004.2868 [astro-ph.IM]} \BibitemShut
  {NoStop}%
\bibitem [{\citenamefont {{En{\ss}lin}}\ and\ \citenamefont
  {{Frommert}}(2010)}]{2010arXiv1002.2928Ep}%
  \BibitemOpen
  \bibfield  {author} {\bibinfo {author} {\bibfnamefont {T.~A.}\ \bibnamefont
  {{En{\ss}lin}}}\ and\ \bibinfo {author} {\bibfnamefont {M.}~\bibnamefont
  {{Frommert}}},\ }\href@noop {} {\enquote {\bibinfo {title} {{Reconstruction
  of signals with unknown spectra in information field theory with parameter
  uncertainty}},}\ } (\bibinfo {year} {2010}),\ \Eprint
  {http://arxiv.org/abs/1002.2928} {arXiv:1002.2928} \BibitemShut {NoStop}%
\bibitem [{\citenamefont {{Jaynes}}(2003)}]{2003prth.book.....J}%
  \BibitemOpen
  \bibfield  {author} {\bibinfo {author} {\bibfnamefont {E.~T.}\ \bibnamefont
  {{Jaynes}}},\ }\href@noop {} {\emph {\bibinfo {title} {Probability Theory, by
  E.~T.~Jaynes and Edited by G.~Larry Bretthorst, pp.~758.~ISBN
  0521592712.~Cambridge, UK: Cambridge University Press, June 2003.}}},\ edited
  by\ \bibinfo {editor} {\bibnamefont {{Bretthorst, G.~L.}}}\ (\bibinfo {year}
  {2003})\BibitemShut {NoStop}%
\bibitem [{\citenamefont {{Kullback}}\ and\ \citenamefont
  {{Leibler}}(1951)}]{Kullback1951}%
  \BibitemOpen
  \bibfield  {author} {\bibinfo {author} {\bibfnamefont {S.}~\bibnamefont
  {{Kullback}}}\ and\ \bibinfo {author} {\bibfnamefont {R.}~\bibnamefont
  {{Leibler}}},\ }\href {\doibase doi:10.1214/aoms/1177729694.} {\bibfield
  {journal} {\bibinfo  {journal} {Annals of Mathematical Statistics}\ }\textbf
  {\bibinfo {volume} {22 (1)}},\ \bibinfo {pages} {79} (\bibinfo {year}
  {1951})}\BibitemShut {NoStop}%
\bibitem [{\citenamefont {{Shannon}}(1948)}]{Shannon1948}%
  \BibitemOpen
  \bibfield  {author} {\bibinfo {author} {\bibfnamefont {C.~E.}\ \bibnamefont
  {{Shannon}}},\ }\href@noop {} {\bibfield  {journal} {\bibinfo  {journal}
  {Bell System Technical Journal}\ }\textbf {\bibinfo {volume} {27}},\ \bibinfo
  {pages} {379} (\bibinfo {year} {1948})}\BibitemShut {NoStop}%
\bibitem [{\citenamefont {{Hubble}}(1934)}]{1934ApJ....79....8H}%
  \BibitemOpen
  \bibfield  {author} {\bibinfo {author} {\bibfnamefont {E.}~\bibnamefont
  {{Hubble}}},\ }\href {\doibase 10.1086/143517} {\bibfield  {journal}
  {\bibinfo  {journal} {\apj}\ }\textbf {\bibinfo {volume} {79}},\ \bibinfo
  {pages} {8} (\bibinfo {year} {1934})}\BibitemShut {NoStop}%
\bibitem [{\citenamefont {{Kitaura}}\ \emph {et~al.}(2009)\citenamefont
  {{Kitaura}}, \citenamefont {{Jasche}}, \citenamefont {{Li}}, \citenamefont
  {{En{\ss}lin}}, \citenamefont {{Metcalf}}, \citenamefont {{Wandelt}},
  \citenamefont {{Lemson}},\ and\ \citenamefont
  {{White}}}]{2009MNRAS.400..183K}%
  \BibitemOpen
  \bibfield  {author} {\bibinfo {author} {\bibfnamefont {F.~S.}\ \bibnamefont
  {{Kitaura}}}, \bibinfo {author} {\bibfnamefont {J.}~\bibnamefont {{Jasche}}},
  \bibinfo {author} {\bibfnamefont {C.}~\bibnamefont {{Li}}}, \bibinfo {author}
  {\bibfnamefont {T.~A.}\ \bibnamefont {{En{\ss}lin}}}, \bibinfo {author}
  {\bibfnamefont {R.~B.}\ \bibnamefont {{Metcalf}}}, \bibinfo {author}
  {\bibfnamefont {B.~D.}\ \bibnamefont {{Wandelt}}}, \bibinfo {author}
  {\bibfnamefont {G.}~\bibnamefont {{Lemson}}}, \ and\ \bibinfo {author}
  {\bibfnamefont {S.~D.~M.}\ \bibnamefont {{White}}},\ }\href {\doibase
  10.1111/j.1365-2966.2009.15470.x} {\bibfield  {journal} {\bibinfo  {journal}
  {\mnras}\ }\textbf {\bibinfo {volume} {400}},\ \bibinfo {pages} {183}
  (\bibinfo {year} {2009})},\ \Eprint {http://arxiv.org/abs/0906.3978}
  {arXiv:0906.3978} \BibitemShut {NoStop}%
\bibitem [{\citenamefont {{Layzer}}(1956)}]{Layzer1956}%
  \BibitemOpen
  \bibfield  {author} {\bibinfo {author} {\bibfnamefont {D.}~\bibnamefont
  {{Layzer}}},\ }\href {\doibase 10.1086/107366} {\bibfield  {journal}
  {\bibinfo  {journal} {\aj}\ }\textbf {\bibinfo {volume} {61}},\ \bibinfo
  {pages} {383} (\bibinfo {year} {1956})}\BibitemShut {NoStop}%
\bibitem [{\citenamefont {{Coles}}\ and\ \citenamefont
  {{Jones}}(1991)}]{1991MNRAS.248....1C}%
  \BibitemOpen
  \bibfield  {author} {\bibinfo {author} {\bibfnamefont {P.}~\bibnamefont
  {{Coles}}}\ and\ \bibinfo {author} {\bibfnamefont {B.}~\bibnamefont
  {{Jones}}},\ }\href@noop {} {\bibfield  {journal} {\bibinfo  {journal}
  {\mnras}\ }\textbf {\bibinfo {volume} {248}},\ \bibinfo {pages} {1} (\bibinfo
  {year} {1991})}\BibitemShut {NoStop}%
\bibitem [{\citenamefont {{Sheth}}(1995)}]{1995MNRAS.277..933S}%
  \BibitemOpen
  \bibfield  {author} {\bibinfo {author} {\bibfnamefont {R.~K.}\ \bibnamefont
  {{Sheth}}},\ }\href@noop {} {\bibfield  {journal} {\bibinfo  {journal}
  {\mnras}\ }\textbf {\bibinfo {volume} {277}},\ \bibinfo {pages} {933}
  (\bibinfo {year} {1995})},\ \Eprint {http://arxiv.org/abs/astro-ph/9511096}
  {astro-ph/9511096} \BibitemShut {NoStop}%
\bibitem [{\citenamefont {{Kayo}}\ \emph {et~al.}(2001)\citenamefont {{Kayo}},
  \citenamefont {{Taruya}},\ and\ \citenamefont {{Suto}}}]{Kayo2001}%
  \BibitemOpen
  \bibfield  {author} {\bibinfo {author} {\bibfnamefont {I.}~\bibnamefont
  {{Kayo}}}, \bibinfo {author} {\bibfnamefont {A.}~\bibnamefont {{Taruya}}}, \
  and\ \bibinfo {author} {\bibfnamefont {Y.}~\bibnamefont {{Suto}}},\ }\href
  {\doibase 10.1086/323227} {\bibfield  {journal} {\bibinfo  {journal} {\apj}\
  }\textbf {\bibinfo {volume} {561}},\ \bibinfo {pages} {22} (\bibinfo {year}
  {2001})},\ \Eprint {http://arxiv.org/abs/arXiv:astro-ph/0105218}
  {arXiv:astro-ph/0105218} \BibitemShut {NoStop}%
\bibitem [{\citenamefont {{Vio}}\ \emph {et~al.}(2001)\citenamefont {{Vio}},
  \citenamefont {{Andreani}},\ and\ \citenamefont
  {{Wamsteker}}}]{2001PASP..113.1009V}%
  \BibitemOpen
  \bibfield  {author} {\bibinfo {author} {\bibfnamefont {R.}~\bibnamefont
  {{Vio}}}, \bibinfo {author} {\bibfnamefont {P.}~\bibnamefont {{Andreani}}}, \
  and\ \bibinfo {author} {\bibfnamefont {W.}~\bibnamefont {{Wamsteker}}},\
  }\href@noop {} {\bibfield  {journal} {\bibinfo  {journal} {\pasp}\ }\textbf
  {\bibinfo {volume} {113}},\ \bibinfo {pages} {1009} (\bibinfo {year}
  {2001})},\ \Eprint {http://arxiv.org/abs/arXiv:astro-ph/0105107}
  {arXiv:astro-ph/0105107} \BibitemShut {NoStop}%
\bibitem [{\citenamefont {{Neyrinck}}\ \emph {et~al.}(2009)\citenamefont
  {{Neyrinck}}, \citenamefont {{Szapudi}},\ and\ \citenamefont
  {{Szalay}}}]{2009ApJ...698L..90N}%
  \BibitemOpen
  \bibfield  {author} {\bibinfo {author} {\bibfnamefont {M.~C.}\ \bibnamefont
  {{Neyrinck}}}, \bibinfo {author} {\bibfnamefont {I.}~\bibnamefont
  {{Szapudi}}}, \ and\ \bibinfo {author} {\bibfnamefont {A.~S.}\ \bibnamefont
  {{Szalay}}},\ }\href {\doibase 10.1088/0004-637X/698/2/L90} {\bibfield
  {journal} {\bibinfo  {journal} {\apjl}\ }\textbf {\bibinfo {volume} {698}},\
  \bibinfo {pages} {L90} (\bibinfo {year} {2009})},\ \Eprint
  {http://arxiv.org/abs/0903.4693} {arXiv:0903.4693} \BibitemShut {NoStop}%
\bibitem [{\citenamefont {{Dorn}}\ \emph {et~al.}(2015)\citenamefont {{Dorn}},
  \citenamefont {{En{\ss}lin}}, \citenamefont {{Greiner}}, \citenamefont
  {{Selig}},\ and\ \citenamefont {{Boehm}}}]{2015PhRvE..91a3311D}%
  \BibitemOpen
  \bibfield  {author} {\bibinfo {author} {\bibfnamefont {S.}~\bibnamefont
  {{Dorn}}}, \bibinfo {author} {\bibfnamefont {T.~A.}\ \bibnamefont
  {{En{\ss}lin}}}, \bibinfo {author} {\bibfnamefont {M.}~\bibnamefont
  {{Greiner}}}, \bibinfo {author} {\bibfnamefont {M.}~\bibnamefont {{Selig}}},
  \ and\ \bibinfo {author} {\bibfnamefont {V.}~\bibnamefont {{Boehm}}},\ }\href
  {\doibase 10.1103/PhysRevE.91.013311} {\bibfield  {journal} {\bibinfo
  {journal} {\pre}\ }\textbf {\bibinfo {volume} {91}},\ \bibinfo {eid} {013311}
  (\bibinfo {year} {2015})},\ \Eprint {http://arxiv.org/abs/1410.6289}
  {arXiv:1410.6289 [physics.data-an]} \BibitemShut {NoStop}%
\bibitem [{\citenamefont {{En{\ss}lin}}\ \emph {et~al.}(2014)\citenamefont
  {{En{\ss}lin}}, \citenamefont {{Junklewitz}}, \citenamefont {{Winderling}},
  \citenamefont {{Greiner}},\ and\ \citenamefont
  {{Selig}}}]{2014PhRvE..90d3301E}%
  \BibitemOpen
  \bibfield  {author} {\bibinfo {author} {\bibfnamefont {T.~A.}\ \bibnamefont
  {{En{\ss}lin}}}, \bibinfo {author} {\bibfnamefont {H.}~\bibnamefont
  {{Junklewitz}}}, \bibinfo {author} {\bibfnamefont {L.}~\bibnamefont
  {{Winderling}}}, \bibinfo {author} {\bibfnamefont {M.}~\bibnamefont
  {{Greiner}}}, \ and\ \bibinfo {author} {\bibfnamefont {M.}~\bibnamefont
  {{Selig}}},\ }\href {\doibase 10.1103/PhysRevE.90.043301} {\bibfield
  {journal} {\bibinfo  {journal} {\pre}\ }\textbf {\bibinfo {volume} {90}},\
  \bibinfo {eid} {043301} (\bibinfo {year} {2014})},\ \Eprint
  {http://arxiv.org/abs/1312.1349} {arXiv:1312.1349 [astro-ph.IM]} \BibitemShut
  {NoStop}%
\bibitem [{\citenamefont {Campbell}(1897)}]{campbell1897law}%
  \BibitemOpen
  \bibfield  {author} {\bibinfo {author} {\bibfnamefont {J.~E.}\ \bibnamefont
  {Campbell}},\ }\href@noop {} {\bibfield  {journal} {\bibinfo  {journal}
  {Proceedings of the London Mathematical Society}\ }\textbf {\bibinfo {volume}
  {1}},\ \bibinfo {pages} {14} (\bibinfo {year} {1897})}\BibitemShut {NoStop}%
\end{thebibliography}%

\appendix

\section{Facilitating Calculations with Operators\label{sec:Facilitating-Calculations-with}}

By making use of our operator formalism we are able to quickly calculate
expectation values of products of exponentials and polynomials like
those we encountered in chapter \ref{sec:Self-Calibrating}. We calculate the Gibbs free energy from the Hamiltonian we got at
the end of chapter \ref{sec:Self-Calibrating}. Combining the equations
(\ref{eq:Gibbs-formula}) and (\ref{eq:Hamiltonian}) yields
\begin{align}
 & G(m,D)=G(\left(\begin{array}{c}
m_{r}\\
m_{a}
\end{array}\right),\left(\begin{array}{cc}
D_{rr} & D_{ra}\\
D_{ar} & D_{aa}
\end{array}\right))\nonumber \\
 & \widehat{=}\left\langle \text{ln}\left(\tilde{P}(s|d)\right)\right\rangle _{\mathscr{G}(s-m,D)}+\nonumber \\
 & \ \left\langle \frac{1}{2}s^{\dagger}S^{-1}s+\frac{1}{2}\left(d-re^{a}\right)^{\dagger}N^{-1}\left(d-re^{a}\right)\right\rangle _{\mathscr{G}(s-m,D)}\nonumber \\
 & \widehat{=}\bigg\langle -\frac{1}{2}(s-m)^{\dagger}D^{-1}(s-m)\nonumber\\
 & \hspace{0.55cm} +\mbox{ln}\left(\left|2\pi D\right|^{-\frac{1}{2}}\right)\bigg\rangle_{\mathscr{G}(s-m,D)}\nonumber \\
 & \ +\left\langle \frac{1}{2}s^{\dagger}S^{-1}s\right\rangle _{\mathscr{G}(s-m,D)}-\left\langle d^{\dagger}N^{-1}re^{a}\right\rangle _{\mathscr{G}(s-m,D)}\nonumber \\
 & \ +\left\langle \frac{1}{2}\left(re^{a}\right)^{\dagger}N^{-1}re^{a}\right\rangle _{\mathscr{G}(s-m,D)}\ .\label{eq:Gibbs energy calibration}
\end{align}
Here ``$\widehat{=}$'' denotes equality up to irrelevant constants,
which are constants that do not depend on $m$ or $D$. 

Following the formalism introduced in \ref{sec:Formulating-Gaussian-Averages}
we replace 
\begin{align}
a\leftrightarrow\Phi^{a} & =D_{ax}\frac{\delta}{\delta m_{x}}+m_{a}\\
r\leftrightarrow\Phi^{r} & =D_{rx}\frac{\delta}{\delta m_{x}}+m_{r}\\
s\leftrightarrow\Phi & =D\frac{\delta}{\delta m}+m
\end{align}

We now evaluate the terms of equation (\ref{eq:Gibbs energy calibration})
one by one. The first two terms are simply second moments of a Gaussian
distribution and thus the calculation can easily be done by hand.
For illustration
we use our formalism anyway and focus on the second term:
\begin{align}
\left\langle \frac{1}{2}s^{\dagger}S^{-1}s\right\rangle _{\mathscr{G}(s-m,D)} &= \frac{1}{2}\int\mbox{d}i\mbox{d}j\,\left\langle s_{i}S_{ij}^{-1}s_{j}\right\rangle _{\mathscr{G}(s-m,D)}\nonumber \\
&=  \frac{1}{2}\int\mbox{d}i\mbox{d}j\,\Phi_{i}S_{ij}^{-1}\Phi_{j}1
\end{align}
We separate 
\begin{align}
\Phi_{t}^{x}=(D\frac{\delta}{\delta m_{x}})_{t}+\left(m_{x}\right)_{t}=c_{t}^{x}+b_{t}^{x}
\end{align}
with 
\begin{align} 
c_{t}^{x}=(D\frac{\delta}{\delta m})_{t}=\int\mbox{d}v\,D_{tv}\frac{\delta}{\left(\delta m_{x}\right)_{v}}\ ,\ b_{t}^{x}=\left(m_{x}\right)_{t}
\end{align}
 where $x$ labels ``$a$'',
``$r$'', or ``'' and arrive at the commutation relations
\begin{align}
[b_{i}^{x},b_{j}^{y}]=[c_{i}^{x},c_{j}^{y}] & =0\\
{}[c_{i}^{x},b_{j}^{y}] & =\left(D_{xy}\right)_{ij}
\end{align}
Our goal is to get the annihilation operators to the right hand side because
they cancel.
Doing so one gets 
\begin{align}
 &\frac{1}{2}\int\mbox{d}i\mbox{d}jS_{ij}^{-1}\Phi_{i}\Phi_{j}1 \nonumber\\
 & =\frac{1}{2}\int\mbox{d}i\mbox{d}jS_{ij}^{-1}\left(c_{i}+b_{i}\right)\left(c_{j}+b_{j}\right)1\nonumber \\
 & =\frac{1}{2}\int\mbox{d}i\mbox{d}jS_{ij}^{-1}\left([c_{i},b_{j}]+b_{i}b_{j}\right)1\nonumber \\
 & =\frac{1}{2}m^{\dagger}S^{-1}m+\frac{1}{2}\mbox{tr}\left(S^{-1}D\right)\ .
\end{align}
We proceed with the third term 
\begin{align}\chapter
 & \left\langle d^{\dagger}N^{-1}re^{a}\right\rangle _{\mathscr{G}(s-m,D)} =d^{\dagger}N^{-1}\Phi^{r}e^{\Phi^{a}}1
\end{align}
To simplify $\Phi^{r}e^{\Phi^{a}}1$ we apply the BCH formula:
\begin{align}
\left(\Phi^{r}e^{\Phi^{a}}1\right)_{j} & =\int\mbox{d}i\,\left(b_{ji}^{r}+c_{ji}^{r}\right)e^{c_{i}^{a}+b_{i}^{a}}1\nonumber \\
 & =\int\mbox{d}i\,\left(b_{ji}^{r}+c_{ji}^{r}\right)e^{b_{i}^{a}+\frac{1}{2}\left(D_{aa}\right)_{ii}}e^{c_{i}^{a}}1
\end{align}
To exchange $c_{ji}^{r}$ and $e^{b_{i}^{a}}$ we use the fact that
$[c_{ji}^{r},\_]$ has the algebraic properties of a derivation, thus
\begin{align}
\Phi^{r}e^{\Phi^{a}}1 & =\int\mbox{d}i\,\left(b_{ji}^{r}+\left(D_{ra}\right)_{(ji)i}\right)e^{b_{i}^{a}+\frac{1}{2}\left(\widehat{D_{aa}}\right)_{i}}e^{c_{i}^{a}}1\nonumber \\
 & =\int\mbox{d}i\,\left(\left(m_{r}\right)_{ji}+\left(D_{ra}\right)_{(ji)i}\right)\left(e^{m_{a}+\frac{1}{2}\widehat{D_{aa}}}\right)_{i}\ .
\end{align}
With $\widehat{D_{aa}}$ we denote the diagonal of the operator $D_{aa}$. We used that if we Taylor expand $e^{c_{i}^{a}}$ only the first
term will contribute since all terms containing $c$ cancel with the $1$.

The last term of equation (\ref{eq:Gibbs energy calibration}) is
\begin{align}
 & \left\langle \frac{1}{2}\left(re^{a}\right)^{\dagger}N^{-1}re^{a}\right\rangle\ .
\end{align}
Translating this into operator language we arrive at
\begin{align}
 &\left\langle \frac{1}{2}\left(re^{a}\right)^{\dagger}N^{-1}re^{a}\right\rangle\nonumber\\
 & =e^{\Phi_{a}^{\dagger}}\Phi_{r}^{\dagger}N^{-1}\Phi_{r}e^{\Phi_{a}}1\nonumber \\
 & =\int\mbox{d}i\mbox{d}j\mbox{d}k\mbox{d}l\,\left(e^{\Phi_{a}}\right)_{l}\left(\Phi_{r}\right)_{kl}\left(N^{-1}\right)_{kj}\left(\Phi_{r}\right)_{ji}\left(e^{\Phi_{a}}\right)_{i}1\ .
\end{align}
First we separate the exponentials with the BCH formula as we have done with the previous term and get
\begin{align}
 & \left(N^{-1}\right)_{kj}\left(e^{\Phi_{a}}\right)_{l}\left(e^{\Phi_{a}}\right)_{i}\left(\Phi_{r}\right)_{kl}\left(\Phi_{r}\right)_{ji}1\nonumber \\
= & \left(N^{-1}\right)_{kj}e^{b_{l}^{a}+\frac{1}{2}\left(D_{aa}\right)_{ll}}e^{c_{l}^{a}}e^{b_{i}^{a}+\frac{1}{2}\left(D_{aa}\right)_{ii}}e^{c_{i}^{a}}\nonumber\\
& \quad\left(c_{kl}^{r}b_{ji}^{r}+b_{kl}^{r}b_{ji}^{r}\right)1\nonumber \\
= & \left(N^{-1}\right)_{kj}e^{b_{l}^{a}+\frac{1}{2}\left(D_{aa}\right)_{ll}}e^{c_{l}^{a}}e^{b_{i}^{a}+\frac{1}{2}\left(D_{aa}\right)_{ii}} e^{c_{i}^{a}}\nonumber\\
& \quad\left(\left(D_{rr}\right)_{\left(kl\right)\left(ji\right)}+b_{kl}^{r}b_{ji}^{r}\right)1\nonumber \\
= & \left(N^{-1}\right)_{kj}e^{b_{l}^{a}+\frac{1}{2}\left(D_{aa}\right)_{ll}}e^{b_{i}^{a}+\frac{1}{2}\left(D_{aa}\right)_{ii}}e^{\left(D_{aa}\right)_{li}}e^{c_{l}^{a}}e^{c_{i}^{a}}\nonumber\\
& \quad\left(\left(D_{rr}\right)_{\left(kl\right)\left(ji\right)}+b_{kl}^{r}b_{ji}^{r}\right)1\nonumber\\
=&  \left(N^{-1}\right)_{kj}e^{b_{l}^{a}+\frac{1}{2}\left(D_{aa}\right)_{ll}}e^{b_{i}^{a}+\frac{1}{2}\left(D_{aa}\right)_{ii}}e^{\left(D_{aa}\right)_{li}}e^{c_{l}^{a}}\nonumber \\
 & \quad\left(\left(D_{rr}\right)_{\left(kl\right)\left(ji\right)}+\left(b_{kl}^{r}+\left(D_{ra}\right)_{\left(kl\right)i}\right)e^{c_{i}^{a}}b_{ji}^{r}\right)1\nonumber \\
=&  \left(N^{-1}\right)_{kj}e^{b_{l}^{a}+\frac{1}{2}\left(D_{aa}\right)_{ll}}e^{b_{i}^{a}+\frac{1}{2}\left(D_{aa}\right)_{ii}}e^{\left(D_{aa}\right)_{li}}\nonumber \\
 & \quad\Big(\left(D_{rr}\right)_{\left(kl\right)\left(ji\right)}+\left(b_{kl}^{r}+\left(D_{ra}\right)_{\left(kl\right)l}+\left(D_{ra}\right)_{\left(kl\right)i}\right)\nonumber \\
 &\quad\left(b_{ji}^{r}+\left(D_{ra}\right)_{\left(ji\right)l}+\left(D_{ra}\right)_{\left(ji\right)i}\right)\Big)1\nonumber \\
=&  \left(N^{-1}\right)_{kj}\left(e^{m_{a}+\frac{1}{2}\widehat{D_{aa}}}\right)_{l}\left(e^{m_{a}+\frac{1}{2}\widehat{D_{aa}}}\right)_{i}\left(e^{D_{aa}}\right)_{li}\nonumber \\
 & \quad\Big(\left(D_{rr}\right)_{\left(kl\right)\left(ji\right)}+\left(\left(m_{r}\right)_{kl}+\left(D_{ra}\right)_{\left(kl\right)l}+\left(D_{ra}\right)_{\left(kl\right)i}\right)\nonumber \\
 &\quad\left(\left(m_{r}\right)_{ji}+\left(D_{ra}\right)_{\left(ji\right)l}+\left(D_{ra}\right)_{\left(ji\right)i}\right)\Big)
\end{align}
For the Gibbs energy we therefore arrive at 
\begin{align}
 & G(m,D)=\nonumber\\
 & -\frac{1}{2}\mbox{tr}\left(1+\mbox{ln}\left(2\pi D\right)\right)+\frac{1}{2}m^{\dagger}S^{-1}m+\frac{1}{2}\mbox{tr}\left(S^{-1}D\right)\nonumber \\
 & -\int\mbox{d}i\mbox{d}j\,\left(d^{\dagger}N^{-1}\right)_{j}\left(\left(m_{r}\right)_{ji}+\left(D_{ra}\right)_{(ji)i}\right)\nonumber\\
 & \quad\left(e^{m_{a}+\frac{1}{2}\widehat{D_{aa}}}\right)_{i}\nonumber \\
 & +\frac{1}{2}\int\mbox{d}i\mbox{d}j\mbox{d}k\mbox{d}l\,\left(e^{m_{a}+\frac{1}{2}\widehat{D_{aa}}}\right)_{l}\nonumber \\
 & \quad\Big(\left(D_{rr}\right)_{\left(kl\right)\left(ji\right)}+\left(\left(m_{r}\right)_{kl}+\left(D_{ra}\right)_{\left(kl\right)l}+\left(D_{ra}\right)_{\left(kl\right)i}\right)\nonumber \\
 &\quad\left(\left(m_{r}\right)_{ji}+\left(D_{ra}\right)_{\left(ji\right)l}+\left(D_{ra}\right)_{\left(ji\right)i}\right)\Big)\nonumber \\
 & \quad\left(N^{-1}\right)_{kj}e^{\left(D_{aa}\right)_{li}}\left(e^{m_{a}+\frac{1}{2}\widehat{D_{aa}}}\right)_{i}\ .
\end{align}
We separately compute the derivative for the reconstructed signal
\begin{align}
 & \frac{\delta G(m,D)}{\delta\left(m_{a}\right)_{i}}= \left(A^{-1}m_{a}\right)_{i}\nonumber\\
 & -\int\mbox{d}j\,\left(d^{\dagger}N^{-1}\right)_{j}\left(\left(m_{r}\right)_{ji}+\left(D_{ra}\right)_{(ji)i}\right)\left(e^{m_{a}+\frac{1}{2}\widehat{D_{aa}}}\right)_{i}\nonumber \\
 & +\int\mbox{d}j\mbox{d}k\mbox{d}l\,\left(e^{m_{a}+\frac{1}{2}\widehat{D_{aa}}}\right)_{l}\nonumber \\
 & \quad\Big(\left(D_{rr}\right)_{\left(kl\right)\left(ji\right)}+\left(\left(m_{r}\right)_{kl}+\left(D_{ra}\right)_{\left(kl\right)l}+\left(D_{ra}\right)_{\left(kl\right)i}\right)\nonumber \\
 & \quad\left(\left(m_{r}\right)_{ji}+\left(D_{ra}\right)_{\left(ji\right)l}+\left(D_{ra}\right)_{\left(ji\right)i}\right)\Big)\left(N^{-1}\right)_{kj}\nonumber\\
 & \quad e^{\left(D_{aa}\right)_{li}}\left(e^{m_{a}+\frac{1}{2}\widehat{D_{aa}}}\right)_{i}
\end{align}
and response
\begin{align}
 & \frac{\delta G(m,D)}{\delta\left(m_{r}\right)_{ij}}= \left(R^{-1}m_{r}\right)_{ij}\nonumber\\
 & -\left(d^{\dagger}N^{-1}\right)_{j}\left(e^{m_{a}+\frac{1}{2}\widehat{D_{aa}}}\right)_{i}\nonumber\\
 & +\int\mbox{d}k\mbox{d}l\,\left(e^{m_{a}+\frac{1}{2}\widehat{D_{aa}}}\right)_{l}\nonumber \\
 & \quad\left(\left(m_{r}\right)_{kl}+\left(D_{ra}\right)_{\left(kl\right)l}+\left(D_{ra}\right)_{\left(kl\right)j}\right)\left(N^{-1}\right)_{ki}\nonumber\\
 & \quad e^{\left(D_{aa}\right)_{lj}}\left(e^{m_{a}+\frac{1}{2}\widehat{D_{aa}}}\right)_{j}\ .
\end{align}
To finalize the derivation we take the second derivative of the Gibbs
free energy which will give us an estimate for $D^{-1}$. Via the
relationship
\[
D^{-1}=\frac{\delta^{2}G(m,D)}{\delta m\delta m^{\dagger}}
\]
We compute
\begin{align}
 & \frac{\delta^{2}G(m,D)}{\delta(m_{a})_{i}\delta(m_{a}^{\dagger})_{l}}=A_{il}^{-1}\nonumber\\
 & -\int\mbox{d}j\,\left(d^{\dagger}N^{-1}\right)_{j}\left(\left(m_{r}\right)_{ji}+\left(D_{ra}\right)_{(ji)i}\right)\nonumber \\
 & \quad\left(e^{m_{a}+\frac{1}{2}\widehat{D_{aa}}}\right)_{i}\delta_{il}\nonumber \\
 & +\int\mbox{d}j\mbox{d}k\mbox{d}n\,\left(e^{m_{a}+\frac{1}{2}\widehat{D_{aa}}}\right)_{n}\Big(\left(D_{rr}\right)_{\left(kn\right)\left(ji\right)}\nonumber \\
 & \quad+\left(\left(m_{r}\right)_{kn}+\left(D_{ra}\right)_{\left(kn\right)n}+\left(D_{ra}\right)_{\left(kn\right)i}\right)\nonumber \\
 & \quad\left(\left(m_{r}\right)_{ji}+\left(D_{ra}\right)_{\left(ji\right)n}+\left(D_{ra}\right)_{\left(ji\right)i}\right)\Big)\nonumber \\
 & \quad\left(N^{-1}\right)_{kj}e^{\left(D_{aa}\right)_{ni}}\left(e^{m_{a}+\frac{1}{2}\widehat{D_{aa}}}\right)_{i}\left(\delta_{il}+\delta_{nl}\right)
\end{align}
\begin{align}
 & \frac{\delta^{2}G(m,D)}{\delta(m_{a})_{k}\delta(m_{r}^{\dagger})_{ji}}= -\left(d^{\dagger}N^{-1}\right)_{j}\left(e^{m_{a}+\frac{1}{2}\widehat{D_{aa}}}\right)_{i}\delta_{ik}\nonumber\\
 & +\int\mbox{d}m\mbox{d}l\mbox{d}n\,\left(\delta_{kl}+\delta_{kj}\right)\left(e^{m_{a}+\frac{1}{2}\widehat{D_{aa}}}\right)_{l}\nonumber \\
 & \quad\left(\left(m_{r}\right)_{nl}+\left(D_{ra}\right)_{\left(nl\right)l}+\left(D_{ra}\right)_{\left(nl\right)j}\right)\left(N^{-1}\right)_{ni}\nonumber \\
 & \quad e^{\left(D_{aa}\right)_{lj}}\left(e^{m_{a}+\frac{1}{2}\widehat{D_{aa}}}\right)_{j}
\end{align}
\begin{align}
 & \frac{\delta^{2}G(m,D)}{\delta(m_{r})_{kl}\delta(m_{r}^{\dagger})_{ji}}= R^{-1}\nonumber\\
 & +\left(e^{m_{a}+\frac{1}{2}\widehat{D_{aa}}}\right)_{l}\left(N^{-1}\right)_{ki}e^{\left(D_{aa}\right)_{lj}}\left(e^{m_{a}+\frac{1}{2}\widehat{D_{aa}}}\right)_{j}
\end{align}
where with $\delta_{xy}$ we denote the Dirac delta function. 
Now we arrived at a point where we have a fully operational reconstruction
algorithm. By using a minimization technique like gradient descent one
can simultaneously reconstruct the signal field and response operator
for given data $d$.

\end{document}